\documentclass[doublecol]{epl2} 
\usepackage{amsmath}
\usepackage{amssymb}
\usepackage{graphicx}

\title{Plasmon induced transparency in graphene based terahertz metamaterials}
\shorttitle{Plasmon induced transparency in graphene based terahertz metamaterials} 

\author{Koijam Monika Devi\inst{1}\thanks{E-mail: \email{koijam@iitg.ernet.in}} \and Maidul Islam\inst{1} \and Dibakar Roy Chowdhury\inst{2} \and Amarendra K. Sarma\inst{1} \and Gagan Kumar\inst{1}}
\shortauthor{K. Monika Devi \etal}

\institute{                    
  \inst{1}  Department of Physics, Indian Institute of Technology Guwahati, Guwahati– 781039, Assam, India\\
  \inst{2} Mahindra \'Ecole Centrale, Jeedimetla, Hyderabad 500043, Telangana, India}


\abstract{
Plasmon induced transparency (PIT) effect in a terahertz graphene metamaterial is numerically and theoretically analyzed. The proposed metamaterial comprises of a pair of graphene split ring resonators placed alternately on both sides of a graphene strip of nanometer scale. The PIT effect in the graphene metamaterial is studied for different vertical and horizontal configurations. Our results reveal that there is no PIT effect in the graphene metamaterial when the centers of both the split ring resonators and the graphene strip are collinear to each other. This is a noteworthy feature, as the PIT effect does not vanish for similar configuration in a metal-based metamaterial structure. We have further shown that the PIT effect can be tuned by varying the Fermi energy of graphene layer. A theoretical model using the three level plasmonic system is established in order to validate the numerical results.  Our studies could be significant in designing graphene based frequency agile ultra-thin devices for terahertz applications.}

\begin{document}
\maketitle

\section{Introduction}
Electromagnetically induced transparency (EIT) is an interference effect that occurs in a three level atomic system. In EIT, an atom that absorbs a particular light signal is rendered transparent by shining another light signal having nearly the same resonance frequency \cite{garrido2002classical}. In 2008, a novel study done by Zhang and his co-workers revealed that an EIT like phenomenon can occur in plasmonic metamaterials (MMs) \cite{zhang2008plasmon}. The plasmonic analogue of this EIT effect is known as the Plasmon Induced Transparency (PIT) effect in MMs\cite{zhang2008plasmon, papasimakis2008metamaterial,chiam2009analogue,souza2015electromagnetically}. PIT usually occurs as a result of interference between the bright and dark plasmonic modes. The bright mode strongly couples with the incident light while the dark mode couples weakly with the incident light \cite{roy2013ultrafast}. For the PIT effect to occur, both the bright and the dark modes should have similar resonant frequencies \cite{zhang2008plasmon,chowdhury2014orthogonally}. Then, the destructive interference of these modes induces a narrow transparency region in the otherwise absorptive spectrum. Since the detection of the PIT effect, a lot of theoretical as well as experimental studies have been done revealing its potential in various applications such as ultrafast sensing \cite{dong2010enhanced,liu2009planar,amin2013dynamically}, switching applications \cite{amin2013dynamically}, slow light phenomena \cite{yannopapas2009electromagnetically,huang2011subwavelength,wang2015tunable,manjappa2015tailoring}, terahertz modulations\cite{devi2017plasmon} etc.

For most of the metal based MMs, tuning of the PIT effect is achieved by altering the geometrical parameters. Recently, active tuning of the PIT effect has also been achieved by incorporating nonlinear media or semiconductor material in the MM structure \cite{gu2012active,cao2013plasmon}. However, only limited tuning could be achieved in case of the metal structures. The quest for a better and efficient tuning of the PIT effect, led to the study of this effect using graphene based MMs. Graphene provides extreme field confinement and low propagation losses. Its Fermi energy can be varied easily by applying an external gate voltage or through chemical doping\cite{garcia2014graphene,ju2011graphene,huang2016graphene,grigorenko2012graphene, pirruccio2017enhancing}. This has kindled a lot of interest in exploring the efficient dynamic tunabilty of the PIT effect in the graphene MMs in the mid-infrared\cite{shi2013plasmonic,zhu2014chip,cheng2013dynamically,fu2016dynamically,shi2015enhanced,sun2017independently,zhao2015plasmon, chen2017dynamically} as well as terahertz frequency regions. Tuning of slow light through PIT effect in graphene MM structures has been studied through the variation of Fermi energy \cite{cheng2013dynamically,fu2016dynamically,shi2015enhanced}. Metal split ring resonators (SRRs) coupled to a graphene strip has also been found to exhibit the PIT effect \cite{zhao2015plasmon}. Comb like structure \cite{shi2013plasmonic,zhang2014plasmon} as well as dipole-dipole coupling \cite{shang2016realization}, a graphene ring coupled to a graphene strip \cite{zhang2017novel}, intergrated graphene waveguides or arrays\cite{lin2015combined, wen2017dynamically} are some configurations that has been investigated in recent years. In spite of these investigations, none of the study has been focused in exploring the PIT effect in graphene based terahertz MM structures to the best of our knowledge. The field of graphene and terahertz MMs has grown significantly in last few years. Consequently, there is ample scope to explore and understand the PIT effect in graphene based MMs and optimize their performance to actualize the construction of terahertz devices.

In this article, we report a novel graphene MM comprising of a pair of graphene SRRs placed alternately on both sides of the graphene strip. The PIT effect in a terahertz graphene MM structure is numerically and theoretically analyzed. The transmission characteristics of the proposed graphene MM structure are studied for different vertical configurations. As opposed to similar metal MM structure, the PIT effect in the graphene MM vanishes when the centers of both graphene SRRs and the graphene strip are collinear to each other. We have also thoroughly examined the PIT effect in the graphene MM when the SRRs are displaced horizontally w.r.t. the graphene strip. Further, the PIT effect is studied for different values of Fermi energy of the graphene material. Finally, a theoretical model based on the three level plasmonic system is established in order to validate our numerical observations. 

\section{Design of graphene terahertz metamaterial}
\begin{figure}[htbp]
	\onefigure{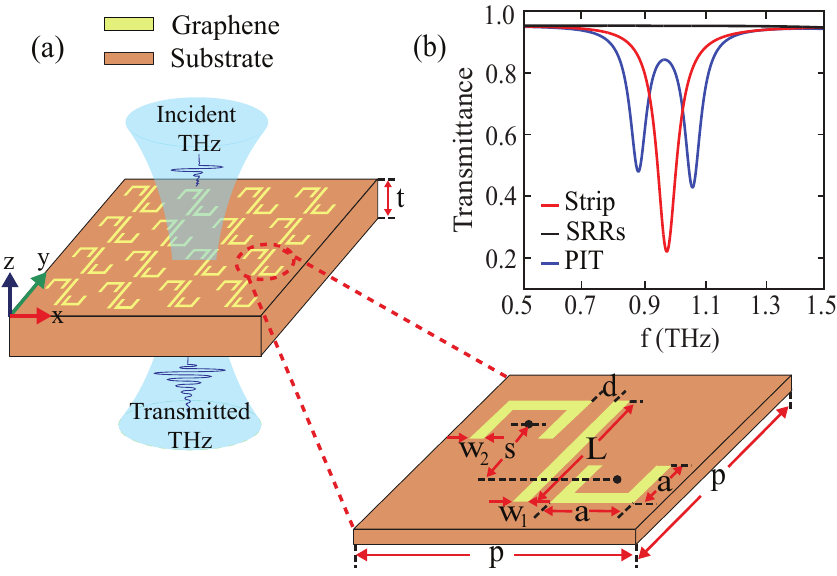}
	\caption{(a) Schematic diagram of the graphene MM structure, (b) Individual transmission profile versus frequency for the graphene strip, graphene SRRs and the graphene PIT MM.}
	\label{fig.1}
\end{figure} The schematic diagram of the graphene MM structure is shown in fig.~\ref{fig.1}(a). The transmission characteristics of the graphene MM structure are investigated using finite  element method based frequency domain solver in the CST Microwave Studio. The MM structure is composed of a strip and a pair of SRRs made of graphene material having a thickness of 1 nm placed on a substrate of thickness $t = 10\  \mu m$, having a dielectric permittivity of 2.4. The graphene strip and the graphene SRRs are designed such that they have similar resonance frequency, with very little deviation, upon excitation by the incident light. The length and width of the graphene strip is $L = 19\ \mu m$ and $w_{1} = 1.1\ \mu m$, respectively. The outer dimensions of each graphene SRR is $a \times a = 6.7\ \mu m \times 6.7\ \mu m $ and width is $w_{2} = 0.9\ \mu m$. $'d'$ is the distance between the graphene strip and the graphene SRRs while $'s'$ is the distance between the centers of the two graphene SRRs. The periodicity of the structure is taken as $p = 24\ \mu m$. The MM structure is simulated under unit cell boundary conditions in the x-y plane. An adaptive mesh size of the order of ${\lambda}/{10}$, where $\lambda$ is the wavelength of the incident radiation, is employed. Open boundary conditions is set along the direction of light propagation. The conductivity of graphene in the terahertz frequency range is described by a simplified Drude expression \cite{garcia2014graphene}:
\begin{equation} \label{eq.1}
\sigma(\omega) =\frac{e^{2}E_{f}}{(\pi \hbar)^{2}} \frac{i}{\omega + i \tau^{-1} }  
\end{equation}
where $e$ is the electronic charge and $ E_{f}$ is the Fermi energy and $\tau$ is the intrinsic relaxation time of the graphene material. The transmission results of the graphene strip, graphene SRRs and the hybrid graphene MM structures are shown in fig.~\ref{fig.1}(b). The red line represents the transmittance of the graphene strip while the black traces represent the transmittance of the graphene SRRs. It is evident from the figure that the graphene strip is directly excited by the incident radiation whereas the fundamental mode of the pair of graphene SRRs is not excited by the incident radiation. Hence, the graphene strip behaves as the bright mode while the graphene SRRs behave as the dark mode. When these graphene SRRs are kept in the vicinity of the graphene strip, they couple through the induced electric field of the graphene strip. This coupling causes a PIT effect in the graphene MM structure, which is represented by the blue line in fig.~\ref{fig.1}(b).

\section{Plasmon induced transparency: Numerical simulations}
\begin{figure}[htbp]
	\onefigure{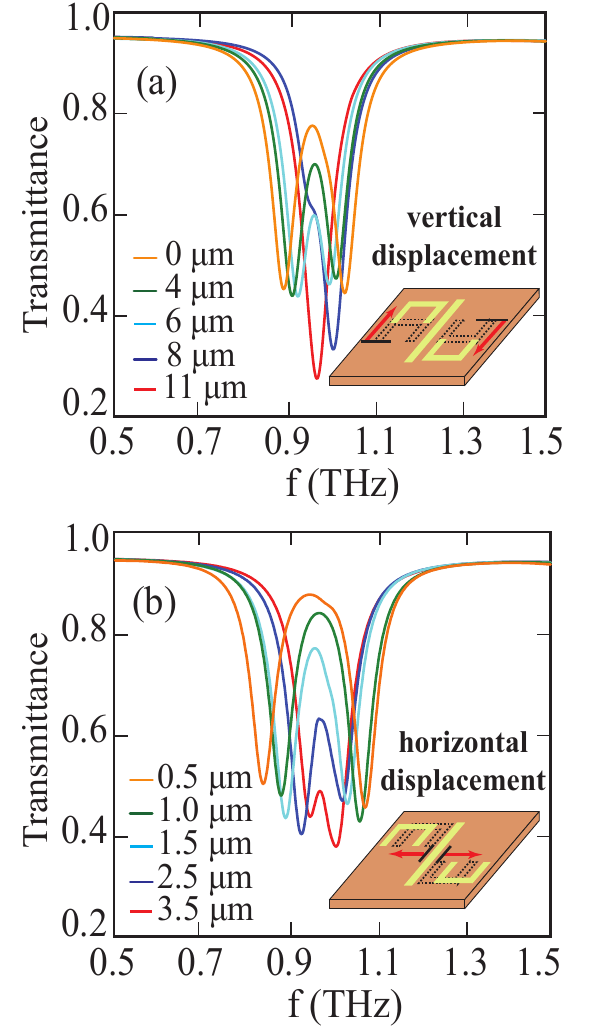}
	\caption{ Numerically simulated transmittance of the graphene MM structure for (a) different values of vertical distance $'s'$ by fixing the horizontal distance, $d =1.5\ \mu m$ and (b) different values of the horizontal distance $'d'$ for a fixed value of vertical distance, $s = 11 \ \mu m$.}
	\label{fig.2}
\end{figure}

The transmission characteristics of the proposed MM structure are studied for different vertical and horizontal configurations. Fig.~\ref{fig.2}(a) represents the transmittance of the graphene MM structure for different variations in $'s'$ while $d=1.5\ \mu m$ is fixed. The red line represents the transmittance of the MM structure when $s=0\ \mu m$. The blue line represents transmittance for $s=4\ \mu m$ while the cyan traces signify the transmittance for $s = 6\ \mu m$. The green and the orange traces represent the transmittance of the MM structure for $s = 8\ \mu m$ and $s = 11\ \mu m$, respectively. Simulations reveal that there is no PIT effect in the graphene MM when both the SRRs are collinear to each other i.e. $s = 0\ \mu m$. However, as the graphene SRRs are displaced away from each other in the vertical direction, the graphene MM starts exhibiting the PIT effect. Next, the transmittance of the graphene MM structure for different values of $'d'$ is shown in fig.~\ref{fig.2}(b). In this case, we have assumed $'s'$ to be fixed. The orange line represents the transmittance of the MM structure when $d = 0.5\ \mu m$. The green line represents the transmittance for $d = 1 \ \mu m$ while the cyan traces signify the transmittance for $d = 1.5 \ \mu m$. The blue and the red traces represent the transmittance for $d = 2.5 \ \mu m$ and $d = 3.5 \ \mu m$, respectively. It may be noted that, as d increases, the PIT window becomes narrower due to the reduced coupling between the graphene strip and the graphene SRRs.

\begin{figure}[htbp]
	\onefigure{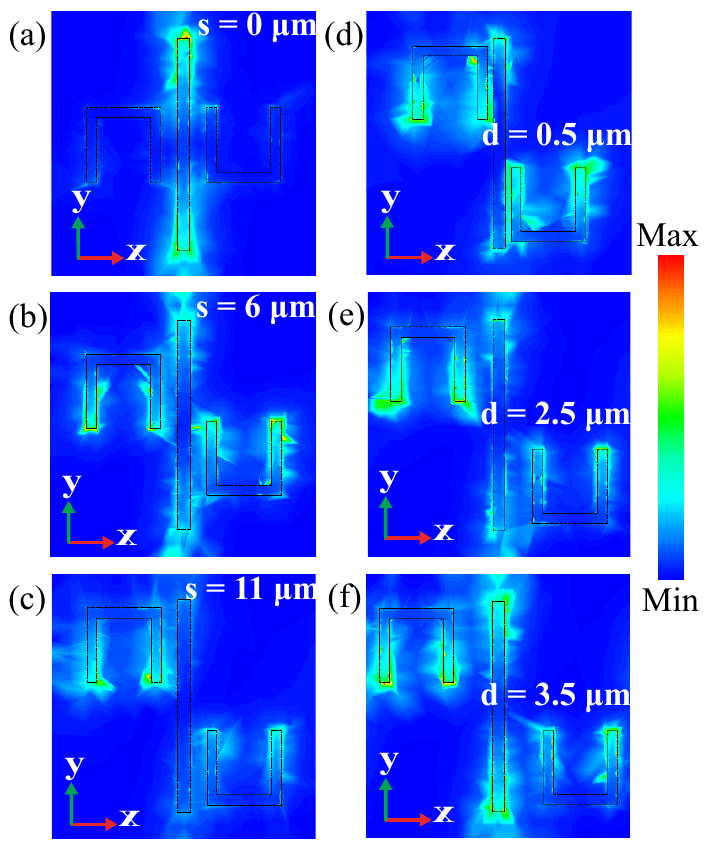}
	\caption{ Electric field profile of the graphene MM structure for different vertical configurations (a) $s = 0 \ \mu m$, (b) $s = 6\ \mu m $, (c) $s = 11 \ \mu m$ and different horizontal configurations  (d) $d = 0.5 \ \mu m$, (e) $ d = 2.5 \ \mu m$ and (f) $ d = 3.5 \ \mu m$. The incident electric field is along the direction of the green arrow. }
	\label{fig.3}
\end{figure}

In order to get insight about the PIT effect in the graphene MM, we study the electric field profiles for various vertical and horizontal configurations. Fig.~\ref{fig.3}(a) to ~\ref{fig.3}(c) represents the electric field profiles of the MM structure for $s = 0 \ \mu m$, $6 \ \mu m$ and $11 \ \mu m$. It is evident from fig.~\ref{fig.3}(a), that the coupling between the graphene strip and the graphene SRRs is negligible for $s = 0\ \mu m$ i.e. when the centers of the graphene SRRs and the graphene strip are collinear. This leads to the vanishing of the PIT effect in the MM structure. This is a very interesting feature as the PIT effect does not vanish for the similar configuration in a metal based MM structure. In metals, when the SRRs are shifted away from the center in the vertical direction, the PIT effect occurs due to inductive coupling via the magnetic field as well as capacitive coupling via the electric field of the middle strip. When the SRRs placed close to the center of the middle strip, the dark mode excitation occurs due to the magnetic field of the strip whereas when the SRRs are close to the edges of the strip, the electric field is responsible for the excitation of dark mode resonance \cite{liu2012electromagnetically}. We believe that in graphene based MM, the dark mode is neither excited by the electric field nor the magnetic field of the graphene strip. However, as the SRRs are displaced away from the center in the opposite direction, the graphene MM structure starts to exhibit the PIT effect.The electric field profile for $s = 6\ \mu m$ is shown in fig.~\ref{fig.3}(b). As the SRRs are further displaced towards the edge of the strip, the PIT window becomes wider due to an increased coupling between the graphene strip and the graphene SRRs. Fig.~\ref{fig.3}(c) represents the electric field profile for $s = 11 \ \mu m$. In this case, the graphene SRRs is close to the edge of the graphene strip and hence the dark mode excitation takes place through the electric field of the graphene strip. The electric field profiles for different horizontal displacements i.e. $d = 0.5\ \mu m$, $2.5 \ \mu m$ and $3.5 \ \mu m$ are also shown in figs.~\ref{fig.3}(d), ~\ref{fig.3}(e) and ~\ref{fig.3}(f), respectively. 
 It is clearly evident from the fig.~\ref{fig.3}(d) to ~\ref{fig.3}(f) that the coupling between the graphene strip and the graphene SRRs is the maximum for $d = 0.5 \ \mu m$.  This configuration corresponds to a wider PIT window (see fig.~\ref{fig.2}(b)). As the graphene SRRs are displaced away from the graphene strip in the horizontal direction, the coupling between them reduces. When $d = 3.5 \ \mu m$, the graphene strip and the graphene SRRs are almost uncoupled. This results in the narrowing of the PIT window and fading of the PIT effect in the MM structure.

 \begin{figure}[htbp]
 	\onefigure{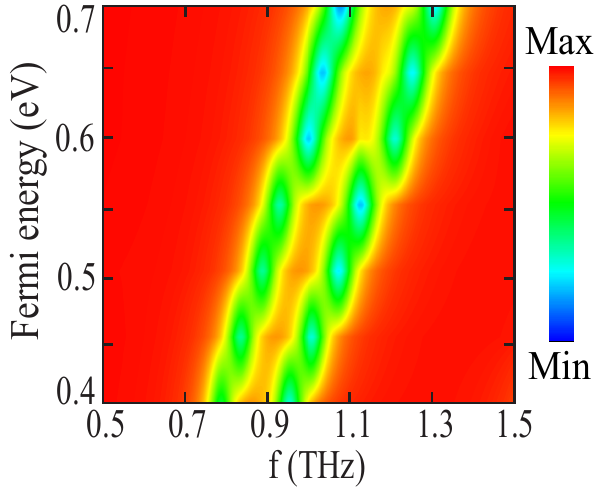}
 	\caption{Contour plot of the transmittance of the graphene MM by varying $E_{f}$ of graphene layer from 0.4 eV to 0.7 eV.}
 	\label{fig.4}
 \end{figure}
Further, to investigate the tunability of the PIT effect in our proposed near field coupled graphene MM, the transmission characteristics were studied by varying the Fermi energy $'{E_{f}}'$ of the graphene layer. Fig.~\ref{fig.4} represents the contour plot of the transmittance of the graphene MM structure for different values of $E_{f}$. When $E_{f} = 0.4$ eV, the PIT peak frequency is observed at $f_R = 0.87$  THz. As the Fermi energy is increased gradually from $0.4$ eV to $0.7$ eV, the PIT peak frequency increases, finally reaching a value of $f_{r} = 1.2$ THz. It is evident from the figure that the PIT window exhibits a blue shift with the increase of the Fermi energy. It may be worthwhile to mention that the resonance frequency of the graphene layer strongly depends upon the Fermi energy by the relation, $f_{r} \propto \sqrt{E_{f}}$ \cite{garcia2014graphene,ju2011graphene,huang2016graphene,grigorenko2012graphene}. This dependence of the peak frequency on the Fermi energy is clearly observed in our numerical results. Hence, an efficient tuning of the PIT effect of the MM structure, without altering the geometrical parameters, can be achieved by tuning the Fermi energy of the graphene layer. This opens up the possibility of realizing reconfigurable or active MMs \cite{roy2011dynamically} based on graphene structures.

\section{Analytical model for PIT effect}
\begin{figure}[htbp]
  	\onefigure{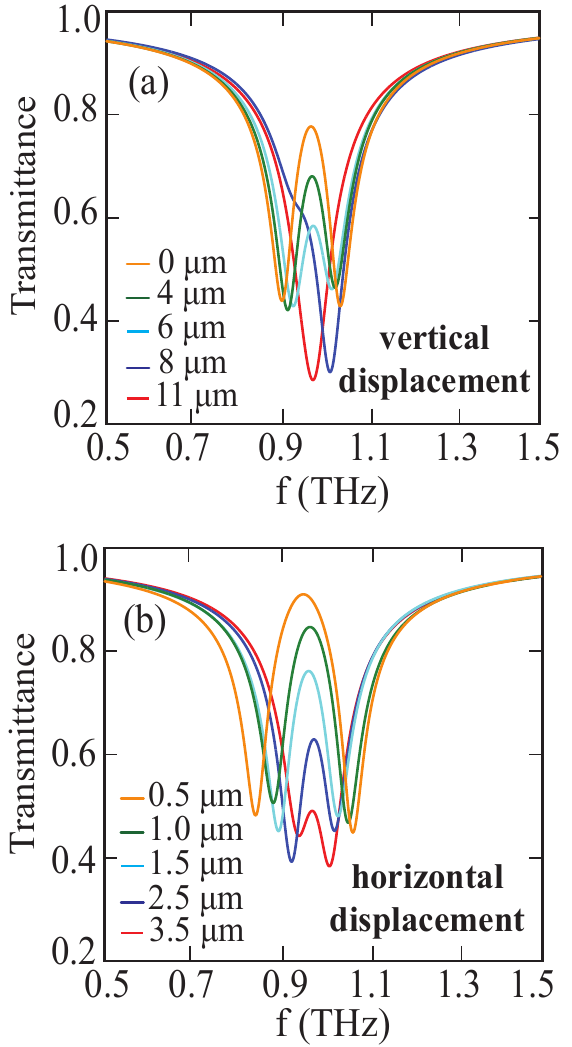}
  	\caption{Theoretically fitted transmittance of the graphene MM structure for (a) different values of  $'s'$  and (b) different values of  $'d'$.}
  	\label{fig.5}
  \end{figure}

In order to better understand the physics of the PIT effect in the graphene MMs, the three level plasmonic model\cite{zhang2008plasmon,cheng2013dynamically,fu2016dynamically} is employed. The model consists of a bright plasmonic state, which strongly couples with the incident field and a dark plasmonic state which does not couple to the incident THz beam. We assume the bright mode and dark mode to have nearly same resonances i.e $\omega_{1}$ and $\omega_{2}$, respectively. The field amplitude of the bright and dark modes can be expressed as 
\begin{equation} \label{eq.2}
\begin{split}
(\omega-\omega_{1}+i \gamma_{1})a_{1}+ \kappa a_{2}= -g_{1}E \\
(\omega-\omega_{2}+i \gamma_{2})a_{2}+ \kappa a_{1}= 0 
\end{split}
\end{equation}
where $\gamma_{1}$, $\gamma_{2}$ are the damping factors of the bright and the dark modes, respectively. $\omega$ is the incident frequency, $\kappa$ is the coupling coefficient between the bright and the dark modes while $g_{1}$ is a parameter describing the coupling between the incident light and the bright resonator. By solving the eq.~(\ref{eq.2}) we obtain
\begin{equation}\label{eq.3}
a_{1}= \frac{g_{1}E(\omega-\omega_{2}+i \gamma_{2})} {\kappa^{2}-(\omega-\omega_{1}+i \gamma_{1})(\omega-\omega_{2}+i \gamma_{2})}
\end{equation} 
Then, the transmission of the graphene MM can be expressed using\cite{cheng2013dynamically}  $t(\omega)=1- |{a_{1}}/{E}|^{2}$ as
follows
\begin{equation}\label{eq.4}
t (\omega)=1- |\frac{g_{1}(\omega-\omega_{2}+i \gamma_{2})} {\kappa^{2}-(\omega-\omega_{1}+i \gamma_{1})(\omega-\omega_{2}+i \gamma_{2})}|^{2}
\end{equation}
The theoretical fitting of the square of the transmission represented by eq.~(\ref{eq.4}) for different values of $'s'$ and $'d'$ are shown in figs.~\ref{fig.5}(a) and ~\ref{fig.5}(b), respectively. The parameters for the theoretical fitting for different values of $'s'$ and $'d'$ are given in table~\ref{tab.1} and in table~\ref{tab.2} , respectively.

\begin{table}[!ht]
	\caption{Parameters for the theoretical fit for various $'s'$.}
	\label{tab.1}
	\begin{tabular}{cccccc}
		\hline \hline 	 
		$s\ (\mu m)$ &$g_{1}$ (THz)&$\gamma_{1}$ (THz)&
		$\gamma_{2}$ (THz) &$\kappa$ (THz) \\
		\hline 
		0 & 0.2 & 0.28 & 0.15 & 0.05  \\
		4 & 0.2& 0.205 & 0.35& 0.25   \\
		6 & 0.2 & 0.15 & 0.36 & 0.31  \\
		8 & 0.2 & 0.12 & 0.37 & 0.36  \\
		11 & 0.2 & 0.105& 0.38 & 0.42 \\
		\hline 	 
	\end{tabular}
\end{table}
\begin{table}[!ht]
	\caption{Parameters for the theoretical fit for various $'d'$.}
	\label{tab.2}
	\begin{tabular}{cccccc}
		\hline \hline
		$d\ (\mu m)$ &$g_{1}$ (THz)&$\gamma_{1}$ (THz)&
		$\gamma_{2}$ (THz) &$\kappa$ (THz) \\
		\hline 
		0.5 & 0.21 & 0.012& 0.45 & 0.7  \\
		1 & 0.21& 0.045 & 0.43& 0.54  \\
		1.5 & 0.21 & 0.11 & 0.36 & 0.42  \\
		2.5 & 0.21 & 0.145 & 0.34 & 0.34 \\
		3.5 & 0.21 & 0.185& 0.33 & 0.26 \\
		\hline 
	\end{tabular}	
\end{table}

The geometrical parameter $g_{1} = 0. 2$ or $0.21$ is same for all the values of $'s'$ and $'d'$, representing a constant free space coupling between the bright mode resonator and electric field of the incident linearly polarized THz beam. $\gamma_{1}$ gradually decreases for variation of $s$ from $0 \ \mu m$ to $11 \ \mu m$ whereas it increases for variation of $d$ from $0.5\ \mu m$ to $3.5\ \mu m$. $\gamma_{2}$ increases for variation of $s$ from $0 \ \mu m$ to $11 \ \mu m$ and decreases for variation of $d$ from $0.5\ \mu m$ to $3.5\ \mu m$. As $'s'$ increases from $0 \ \mu m$ to $11 \ \mu m$, $\kappa$ gradually increases indicating an increase in the coupling between the bright and the dark mode of the graphene MM, which results in the widening of the PIT window (fig.~\ref{fig.5}(a)). $\kappa$ reduces steadily as $'d'$ increases indicating a reduction of coupling between the bright and the dark mode resulting in the narrowing of the PIT window (fig.~\ref{fig.5}(b)). The theoretical transmission results are in good agreement with our numerical results.
\section{Conclusion}
In summary, the plasmon induced transparency effect in a near field coupled metamaterial comprising of a graphene strip and a pair of graphene split ring resonators is numerically and theoretically analyzed. The plasmon induced transparency effect is modulated by varying the vertical displacement as well as horizontal displacement of the split ring resonators w.r.t. the graphene strip. As the graphene SRRs are displaced in the vertical direction, the PIT effect vanishes when the SRRs are collinear to each other. When the graphene SRRs are displaced in the horizontal direction, the PIT window becomes narrower due to reduced coupling between the dark and bright modes. Further, the PIT effect is studied for different values of Fermi energy of the graphene material. The PIT window exhibits a blue shift as the Fermi energy gradually increases from $0.4$ eV to $0.7$ eV. Finally, a theoretical model based on the three level plasmonic system is provided in order to validate our numerical findings. The theoretical results are in good agreement with the numerical results. Our studies may be significant in designing frequency tunable, active and reconfigurable terahertz metamaterials in near future.

\acknowledgments
 The author, GK gratefully acknowledges the financial support from the Board of Research in Nuclear Sciences (BRNS), India (34/20/17/2015/BRNS). Author DRC gratefully acknowledges the financial support from the SERB, Department of Science and Technology, India (EMR/2015/001339). Author KMD would like to thank MHRD, Government of India for a reasearch fellowship.

\end{document}